\documentclass[conference]{IEEEtran}
\IEEEoverridecommandlockouts
\usepackage{cite}
\usepackage{amsmath,amssymb,amsfonts}
\usepackage{algorithmic}
\usepackage{graphicx}
\usepackage{comment}
\usepackage{textcomp}
\usepackage{xcolor}
\def\BibTeX{{\rm B\kern-.05em{\sc i\kern-.025em b}\kern-.08em
    T\kern-.1667em\lower.7ex\hbox{E}\kern-.125emX}}
\begin{document}

\title{How to Define Design in Industrial Control and Automation Software\\
{\footnotesize \textsuperscript{*} The misunderstood concept of design in industrial software applications}
\thanks{I deeply appriciate help and guidance of Prof. Dr. Nam P. Suh in understanding what complixty is all about and how to address it scientificly.}
}

\author{\vspace{1em}}  % Empty author block to suppress errors

\author{\IEEEauthorblockN{Aydin Homay}
\IEEEauthorblockA{\textit{Chair of Industrial Communications} \\
\textit{Technische Universität Dresden}\\
Dresden, Germany \\
https://orcid.org/0000-0002-6425-7468}
}

\maketitle

\begin{abstract}
Design is a fundamental aspect of engineering, enabling the creation of products, systems, and organizations to meet societal and/or business needs. However, the absence of a scientific foundation in design often results in subjective decision-making, reducing both efficiency and innovation. This challenge is particularly evident in the software industry and, by extension, in the domain of industrial control and automation systems (iCAS).

In this study, first we review the existing design definitions within the software industry, challenge prevailing misconceptions about design, review design definition in the field of design theory and address key questions such as: When does design begin? How can design be defined scientifically? What constitutes good design? and the difference between design and design language by relying on advancements in the field of design theory. We also evaluate the distinction between ad-hoc and systematic design approaches, and present arguments on how to balance complementary operational concerns while resolving conflicting evolutionary concerns.
\end{abstract}


\section{Introduction}
G. Booch cites Dijkstra, stating that programming is fundamentally a large-scale application of abstraction, demanding both the precision of a formal mathematician and the practical mindset of a skilled engineer~\cite[pp.~25]{Booch-2007}. When it comes to the formal part of design in software systems, we see that the term "design" is frequently misused and misunderstood, often applied with insufficient precision across various domains. Scholars, professionals, subject matter experts, and students frequently employ this term without providing a clear definition or explanation of its intended meaning. This issue is particularly pronounced in the software industry and, consequently, in the field of industrial control and automation systems (iCAS). 

A substantial body of literature within the iCAS field utilizes the term \textit{design} without dedicating a single line to define what design is all about~\cite{DMCS-1982,John-2010,Vyatkin-2011,Bottcher-2013,Yoong-Roop-Bhatti-2015,Mehta-2015,Dai-Sun-2019,Kshirsagar-2022,Zhou-2022,Dai-2022,Sharma-Zoitl-2023,Bauer-Zoitl-2023,Kotlyarov-etal-2023}. 

For instance, in a work published by \textit{IEEE Transactions on Industrial Informatics}, the author employed the term \textit{design} 98 times within a 12-page paper, yet failed to offer any definition of what design entails~\cite{Lyu-2021}. 

Author of~\cite{Vyatkin-2013} dedicated one section of his paper \textit{"Software Engineering in Industrial Automation: State-of-the-Art Review"} on the topic of \textit{"Software Design"} and used the term "design" 72 times, but did not define what design entails. 

Similarly, authors of "Modeling Control Systems Using IEC 61499"~\cite{Zoitl-Lewis-2014} used the term "design" over a hundred times throughout their book, without any substantive effort to clearly define what design is all about and how to understand it.

To advance any field including design in iCAS, we first must establish a precise and scientific definition that allows us to systematically explore the matter. Without such a clear definition, we cannot form a common understanding of what the problem is that we are trying to solve or how we can improve and advance the current state of the art. i.e. if the meaning and understanding of design vary depending on whom we ask, then we have not defined it scientifically, and we cannot advance it~\cite{Suh-2005}.

Scientific definition is the one that identifies the essential features of a concept, without which it would not be what it is. Founded in realism, scientific definitions follow a genus-differentia structure, categorizing an entity and clearly stating its unique properties, for example, defining water as a substance (genus) with the molecular structure H$_2$O (differentia)\cite{Hibberd-2019}. For instance, consider the first and second laws of thermodynamics. Regardless of whom we inquire, if the individual possesses knowledge of the fundamental principles of classical physics, the response will invariably be consistent.

The remainder of this paper is structured as follows: Section~\ref{state_of_the_art} elucidates the ambiguity of design within the context of software engineering. Section~\ref{design_theory} explores design theory and its implications, while Section~\ref{design_in_engineering} offers an introduction to design within engineering. Section~\ref{design_process} examines the initiation of the design process, and Section~\ref{automation_system_design_concerns} addresses the primary design concerns pertinent to industrial control and automation systems. Section~\ref{design_language} distinguishes between design and design language as the concluding theme, culminating in the conclusion presented in Section~\ref{conclusion}.

\section{State of The Art}
\label{state_of_the_art}
The issue of design that lacks a scientific definition within iCAS is rooted in a clumsy definition of design in the field of software engineering. 

In software engineering, depending on whom we ask, we will get a different answer to what design is, for example:

In 1978, Yourdon and Constantine defined design as an activity that begins once the systems analyst has produced a set of functional requirements for a program or system and ends when the designer has specified the components of the system and the interrelationship between the components~\cite[pp.~16]{Yourdon-1978}.

In 2000, Baldwin et al. defined design as the process of inventing objects, our 'things,' that perform specific functions. These objects, the products of human intelligence and effort, are called 'artifacts'~\cite[pp.~2]{Baldwin-Clark-2000}.

In 2004, C. Larman defined design as a conceptual solution (in software and hardware) that fulfills the requirements, rather than its implementation. For example, a description of a database schema and software objects. Design ideas often exclude low-level or 'obvious' details that are obvious to intended consumers. Ultimately, designs can be implemented, and the implementation (such as code) expresses the true and complete realized design~\cite[pp.~5]{Larman-2005}.

In 2006, I. Sommerville defined design as a creative activity in which you identify software components and their relationships, based on the requirements of a customer. Implementation is the process of realizing the design as a program~\cite[pp.~197]{Sommerville-2016}.

In 2017, R. C. Martin argued that there is no difference between design and architecture as both are part of the same whole. However, he did not define what design is or how to understand architecture~\cite[pp.~4]{R.C.Martin-2017}.

In 2021, Bass et al. defined design for software systems as no different than design in general: It involves making decisions and working with the available materials and skills to satisfy requirements and constraints~\cite[pp.~425]{Bass-Clements-Kazman-2021}.

In 2024, Kazman et al. defined design as a translation from the world of needs (requirements) to the world of solutions, in terms of structures composed of modules, frameworks, and components. A good design is one that satisfies the drivers~\cite[pp.~4]{Cervantes-Kazman-2024}. This definition is the only definition that we found which closely aligns with the definition of design proposed by design theorists during 1980-1990~\cite{Yoshikawa-Uehara-1985, Newell-Simon-1972,Suh-1990}.

Let us end this section by the famous quote of Grady Booch, which says: "All architecture is design, but not all design is architecture."

\section{Design Theory}
\label{design_theory}
Herbert Simon, in "The Sciences of the Artificial" (1969), argued that design could evolve into a structured scientific discipline governed by established theories and practices~\cite{Simon-1996}. He viewed design not merely as an art but as a formalized process that could be systematically understood and applied~\cite{Braha-1998, Simon-1996}. 

Kazman et al., in "Software Architecture in Practice" (2021) and "Designing Software Architectures – A Practical Approach" (2024), reinforce this idea, asserting that design is not an exclusive skill limited to experts. Instead, it can be taught and learned as an accessible and structured discipline. Engineering design primarily involves assembling known design primitives in innovative ways to achieve predictable outcomes, highlighting the need for systematic and repeatable methodologies~\cite{Bass-Clements-Kazman-2021}.  

This aligns with the mid- to late twentieth century shift toward systematic design, which led to its widespread adoption in fields like mechanical engineering. These structured methodologies provided a scientific foundation for repeatable design processes, a transformation that software engineering has not yet fully integrated~\cite{Simon-1996, Yoshikawa-Uehara-1985, Braha-1998, Suh-1990}.  

The next section challenges the misconception that design starts with well-defined requirements, instead exploring its fundamental nature, what constitutes good design, and the distinctions between ad-hoc and systematic approaches.

\subsection{What is Design?}
Braha et al. in their book~\cite[pp.~30]{Braha-1998} state that: Regardless of the particular design domain, "Design", as problem solving, is a natural and ubiquitous human activity that shares fundamental characteristics irrespective of the design domain. Therefore, fundamentally, design can be taught~\cite[pp.~1]{Cervantes-Kazman-2024}, and designers can learn to create optimal designs by applying systematic and scientific principles~\cite{Suh-1990}.

Design begins with the recognition of the need and dissatisfaction with the current state of affairs, which leads to understanding that steps must be taken to address problems (needs)~\cite[pp.30]{Braha-1998}.

Yoshikawa's research from 1981 to 1986~\cite{Tomiyama-1986,Chen-1998} introduced the notion of perceiving design as a mapping process from the problem domain to the solution domain. This conceptual framework was further developed and brought to a culmination by Suh's research~\cite{Suh-2001}.

The definition of design should be revised to begin with business needs rather than requirements because an essential design principle must be considered during the translation of needs into requirements. This principle, known as the independence principle of requirements, is articulated in the definition of functional requirement discussed by Suh in Axiomatic Design~\cite{Suh-2001}. If independence among requirements is not properly established during this translation, any attempt to address coupling in the solution space will be nothing more than a fantasy.

Furthermore, relying solely on gathering requirements through customer interviews and attempting to capture their stated needs makes it nearly impossible to avoid conflicts among requirements. This challenge is supported by Arrow’s Impossibility Theorem~\cite{Arrow-2012}, which demonstrates the inherent difficulty of aggregating multiple preferences into a consistent decision. Furthermore, this approach often results in vague, incomplete, ambiguous, and coupled requirements, as discussed by Wiegers and Beatty~\cite{Wiegers-Beatty-2013}.

\subsection{Good Design}
Historically, many engineers have developed their products (or processes, systems, etc.) through iterative, empirical, and intuitive methods, relying on years of experience, ingenuity, and creativity, often accompanied by a wide variety of trial-and-error and ad-hoc decision-making approaches~\cite{Suh-2001}.

Although experience plays a crucial role in accumulating practical design knowledge, it is not always sufficient. This is especially true when the context of the application changes (e.g., software systems), making experiential knowledge less reliable~\cite{Suh-2001}. Therefore, experience must be complemented by a systematic understanding of design, or vice versa~\cite{Suh-2001} to achieve good design. Good design is a design that effectively satisfies the intended design goals (functional requirements)~\cite{Suh-2001}. 

The philosophy of good design in software (which at the time was based on reusability and changeability) was discussed at the 1968 NATO conference in Garmisch, Germany~\cite{NATO-1968} by McIlroy, Brown, Hopkins, and Warshall, and later culminated by Parnas in his paper "The Criteria to Be Used in Decomposing Systems into Modules"~\cite{Parnas-1972} in 1972 and explicitly Dijkstra in his 1974 paper "On the Role of Scientific Thought"~\cite{Dijkstra-1974}.

Robert C. Martin, in his book "Clean Architecture", defines design quality based on the effort required to fulfill customer needs. A design is considered good if this effort remains consistently low throughout the system's lifespan. In contrast, if the effort increases with each new release, the design is poor~\cite{R.C.Martin-2017}. This definition holds particularly true when reusability and changeability are the primary design concerns. 

L. Bass, P. Clements, and R. Kazman define high-quality software as systems that effectively satisfy key quality attributes such as availability, reliability, maintainability, and security~\cite{Bass-Clements-Kazman-2021}.  

While we fully acknowledge the importance of these attributes, it is essential to recognize that they are typically measured only after the system has been designed and deployed into production, at which point addressing fundamental design flaws may be too late. This does not mean that quality attributes should be ignored; rather, it highlights the need for a mechanism to evaluate design quality as it evolves throughout the development process. A structured approach should enable us to assess these qualities at each stage of design, rather than relying solely on post-deployment evaluations.  

For this reason, we align with Suh's perspective~\cite{Suh-2001}, which defines good design as one that effectively satisfies all functional requirements. This definition is broadly applicable to the evaluation of good design and is also recognized by scholars such as Kazman and Pressman~\cite{Cervantes-Kazman-2024, Pressman-Maxim-2019}. Furthermore, the merit of Suh's definition of good design lies in its provision of two elegant axioms that guide the design process. The independence axiom~\cite[pp.~16]{Suh-2001} ensures that throughout the design process, systems remain uncoupled or decoupled, while the information axiom~\cite[pp.~39]{Suh-2001} seeks to reduce design complexity by minimizing the information content, thereby facilitating the selection of the optimal design among alternative designs. 

\subsection{Ad-hoc and Systematic Design}
\label{adhoc_design_vs_systematic_design}
Unlike ad-hoc design, which relies on empirical methods and is optimized for specific conditions, systematic design aims to establish generalizable principles. It seeks to guide design decisions using predefined and well-founded concepts, such as axioms, theorems, and corollaries, which are based on agreed-upon or proven facts~\cite{Suh-1990, Braha-1998}.

In essence, algorithmic design is an ad-hoc approach that involves the identification or outline of a design process that ultimately results in a design embodiment that meets the intended design goals every time it is employed~\cite{Braha-1998,Suh-2001,Suh-2021-D}. i.e., in algorithmic design, there exist prescribed procedures and these procedures could be exact or heuristic that ultimately result in a design embodiment that satisfies design goals.

For instance, Quick-sort can sort the data in an average time of O(n log n) if element stability is not required and the partitioning algorithm avoids always selecting the largest or smallest element as the pivot. Therefore, algorithmic methods are useful both under specific conditions and during detailed design, where designers address atomic requirements that cannot be further broken down~\cite{Suh-2001}.

The algorithmic approach can be divided into several categories~\cite{Suh-2001, Braha-1998}:

\begin{itemize}
    \item pattern recognition
    \item associative memory
    \item analogy
    \item experimentally based prescription
    \item extrapolation/interpolation
    \item selection based on probability
\end{itemize}

In fact, the algorithmic design paradigm can be utilized as a tool that can be supported and invoked by other design paradigms, that is, ideal design should start with axioms and proceed to algorithms and tools~\cite[pp.~9]{Suh-2001}~\cite[pp.~58]{Braha-1998}.

In the following sections, we will explore the primary issue with design in software engineering: its inherently ad hoc nature. This manifests in two ways—either as heuristic decision-making, where each designer interprets key concepts (e.g., coupling, cohesion, modularization, and encapsulation) differently, or as a rigid algorithmic approach that offers limited flexibility and functions effectively only under specific conditions. Then, to ensure the continuous development and stability of a software product, the only viable strategy is to maximize automated testing, preventing unintended alterations to these predefined conditions. Yet, in principle, we cannot prevent change or modification in requirements by relying on automated testing, as the space and time for designing activity in requirement engineering and design activity in requirement realization (implementation) differ. Therefore, at some point of time change will enter to the system as a result of requirement alternation resulting in low level design re-adjustment and consequently out-dating the design decision that was taken earlier based on widely used software design principles (heuristics and algorithms).  

\subsection{Design in Engineering}
\label{design_in_engineering}
Design is central to engineering, allowing the creation of products, systems, and organizations to meet societal needs. It involves defining problems, synthesizing solutions, and analyzing outcomes to ensure alignment with the intended goals. However, the lack of a scientific foundation in design often leads to subjective decision-making, limiting efficiency and innovation. A structured approach to design, incorporating clear principles and analytical validation, is essential to improve quality, reduce errors, and ensure optimal solutions.

\begin{figure}[ht]
    \centering
    \includegraphics[width=0.3\textwidth]{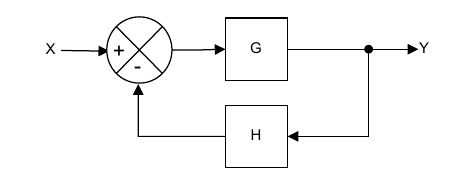}
    \caption{Feedback control loop depicting the design process.).~\cite[pp.~11]{Suh-2001}}
    \label{fig:fundamentals/design_feedback_loop}
\begin{equation}
        \frac{Y}{X} = \frac{G}{1 + GH} \approx \frac{G}{GH} = H^{-1} \quad \text{for } GH \gg 1
        \label{eq:approximation}
\end{equation}
\end{figure}
    
In the~\ref{fig:fundamentals/design_feedback_loop}, \textit{Y} is the desired outcome, and \textit{X} is the input. The gain of the feedback loop should be maximized to converge on an optimal solution efficiently, enhancing the ability to assess the quality of the design. As shown in the figure, the creative process benefits from this analytical validation. When the product of functions \textit{G} and \textit{H} is significantly greater than one, the gain is approximately \textit{1/H}. Without proper analysis, distinguishing between good and bad designs becomes difficult, preventing the rapid generation of an optimal solution. In the absence of clear criteria for the evaluation of designs, effective decision making is compromised~\cite[pp.~7]{Sung-1999}.

\subsection{Design Process}
\label{design_process}
The design process begins with identifying a societal need and translating it into functional requirements. The designer determines these requirements, shaping the design challenge. The concepts are then developed, evaluated, and refined through an iterative feedback process until a satisfactory outcome is reached. An effective designer keeps functional requirements to a minimum while preserving essential functionality to avoid unnecessary complexity. Dependencies between requirements should be avoided unless absolutely necessary to prevent restrictive constraints. While factors like cost, size, and aesthetics influence design decisions, they should not completely dictate them. A well-structured set of functional requirements ensures that the final design effectively fulfills its intended purpose~\cite[pp.~27-28]{Suh-1990}.

Creativity plays a crucial role in the generation of novel solutions. Analogical thinking and interdisciplinary knowledge help designers identify innovative ideas. Many historical innovations resulted from the identification and application of known principles in new ways~\cite[pp.~29-30]{Suh-1990}.

Note that a good designer must be able to operate in the conceptual world of the functional as well as the physical domain~\cite[pp.~28]{Suh-1990}. When translating needs into requirements, it is crucial to maintain a "solution-neutral environment"~\cite[pp.~15]{Suh-2000}. For instance, the designer should ensure that functional requirements remain independent, as interdependent requirements add unnecessary complexity without additional benefits. When the requirements are dependent on each other, they can be combined into one~\cite[pp.~28]{Suh-1990}. Although social needs and customer input play a role in defining functional requirements, they cannot always be determined solely by surveying customers~\cite[pp.~14]{Suh-2000}. This limitation is supported by Arrow's impossibility theorem~\cite{Arrow-2012}.

\begin{figure}[ht]
    \centering
    \includegraphics[width=0.45\textwidth]{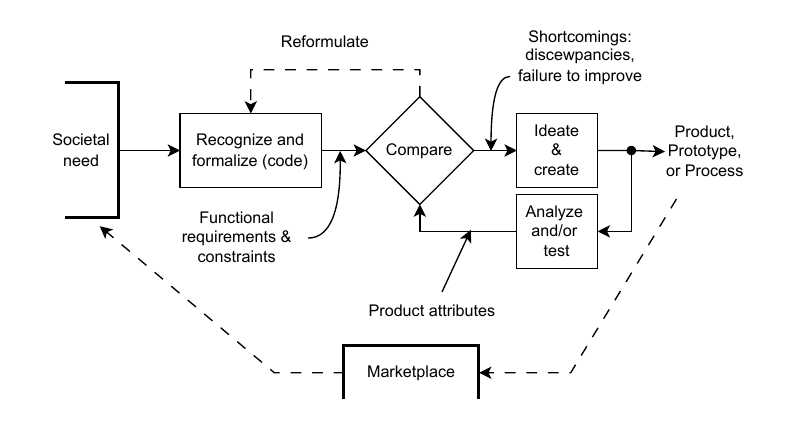}
    \caption{The design loop, more-effective synthesis requires enhanced creativity, more-powerful analysis, and improved decision bases (From Wilson, 1980).~\cite[pp.~27]{Suh-1990}}
    \label{fig:fundamentals:the_design_loop}
\end{figure}

Suh defines the design process as an interplay between "what we want to achieve" and "how we choose to satisfy the need (i.e., what).~\cite[pp.~3]{Suh-2001}". In principle, this definition means the culmination of synthesized solutions (in the form of product, software, processes, or systems) by the appropriate selection of design parameters that satisfy perceived needs through the mapping from functional requirements in the functional domain to design parameters in the physical domain.

\section{Design and Design Concerns in Industrial Automation Systems}
\label{automation_system_design_concerns}
\begin{figure}[h]
    \centering
    \includegraphics[width=1\linewidth]{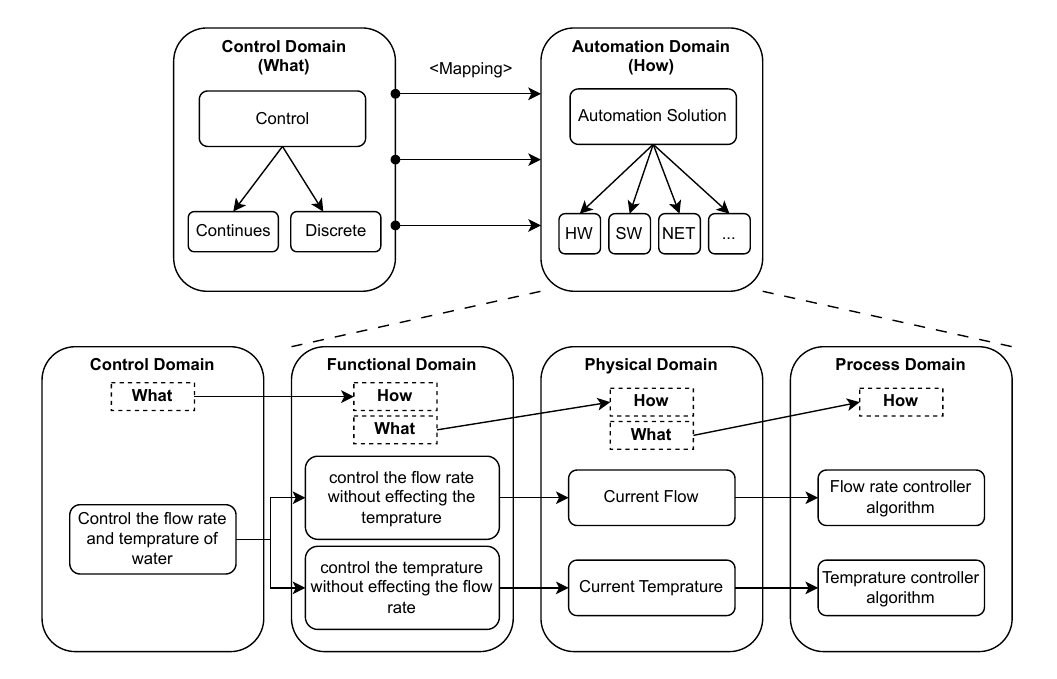}
    \caption{The design process starts from the control domain and continues to automation domain via mapping task.}
    \label{fig:fundamentals:design_definition_for_iCAS}
\end{figure}

Following Suh~\cite{Suh-2001} on axiomatic design, we define \textit{design} in iCAS as a mapping process that translates \textbf{what} the iCAS must achieve and \textbf{how} it must achieve. In other words, the design in iCAS represents the transformation from the control domain to the automation domain. In the control domain, we define the needs of the control process, and in the automation domain, we capture the automation and control solutions (see Fig.~\ref{fig:fundamentals:design_definition_for_iCAS}). A good design must satisfy design concerns and requirements effectively.

Given that design requirements vary across businesses, the objective of this study is to focus on design concerns, which are frequently common across diverse systems. Design concerns, also known as design \textit{Quality Attributes}~\cite{ISO-IEC-25010-2023}, play a crucial role by offering measurable criteria to guide engineers in making informed design decisions. The ISO/IEC 25010 standard~\cite{ISO-IEC-25010-2023} provides a comprehensive list of quality attributes for software systems. However, in this study, we adopt slightly modified terminology and definitions for iCAS-specific concerns, which will be elaborated on in the following sub-sections. To understand which design concern has to be maximized in the design, we need to understand what purpose/goal the system under design is going to serve/achieve.

In this section, we have organized the design considerations of iCAS into two main categories: operational and evolutionary. The objective behind this categorization is to alert the iCAS designer to the phenomenon that as the system designer climbs the automation hierarchy, evolutionary concerns gain prominence, whereas moving down to the field level (level 0) highlights operational concerns. It is critical to note that this does not imply that software systems at lower levels of the pyramid (e.g., Distributed Control Systems, Open Control Systems) do not require evolutionary considerations. In fact, every software system inevitably requires evolutionary capabilities to some extent. However, the designer faces the need to strike a balance between these concerns, contingent on the operational level, thereby determining whether to prioritize operational or evolutionary considerations.

\subsection{Operational Concerns}
Operational (run-time) concerns of a software system refer to the key aspects and challenges that affect its ability to function effectively and efficiently in a production environment. These concerns typically revolve around maintaining reliability, availability, performance, and usability while ensuring that the system meets both user and business expectations. The following are the main operational concerns for an iCAS:

\subsubsection{Reliability} is the ability of a system or component to perform its required functions consistently and predictably under specified conditions for a specified period of time. Reflects the degree to which a system, product, or component delivers agreed or expected functionality during the defined time frame and conditions, as outlined in ISO/IEC 25010:2011. Influenced by the inherent qualities of design, implementation, and contextual factors, as specified in ISO/IEC 16350:2015, reliability is closely related to availability and is often measured using parameters such as the mean time between failures (MTBF)~\cite{ISO-IEC-IEEE-24765-2017,ISO-IEC-25010-2023}.
    
\subsubsection{Predictability} refers to the degree to which the performance of an industrial control and automation system can be accurately predicted. This involves the consistency and stability of the system's behavior over time, allowing operators to anticipate its performance and respond proactively to potential issues.
    
\subsubsection{Availability} relates to the reliability of the system and considers maintainability. To design for availability, it is important to assess the impact of system failures and identify the necessary steps to restore the system's operation or performance to its intended standards~\cite{Stapelberg-2009,Mehta-2015}. 

\subsection{Evolutionary Concerns}
The basis for software evolution management was laid in the 1980s in the computer science domain. Lehman (1980) defined laws of software evolution and, among others, identified that systems are subject to dynamics, causing continuing changes in software that result in increasing complexity~\cite{Lehman-1980}.
Compared to general software development platforms, iCAS did not evolve much. The adaptation of advanced software development concepts like Object-Oriented Programming (OOP), Service-Oriented Architecture (SOA), and Microservices Architecture (MSA) did not fully penetrate the iCAS world from the industry perspective.

\subsubsection{Changeability} based on the definition of \textit{ modifiability} in IEC 25010~\cite{ISO-IEC-25010-2023}, we have defined changeability as the ability of a product to be effectively and efficiently modified without introducing defects or compromising the primary concerns. 
    
\subsubsection{Maintainability} as defined by IEC 25010~\cite{ISO-IEC-25010-2023}, refers to the ability of a product to be modified by its intended maintainers effectively and efficiently. Modifications may include corrections, enhancements, or adaptations to changes in the environment, requirements, or functional specifications. These modifications can be performed by specialized support staff, operational staff, business users, or even end users. Maintainability also includes the installation of updates and upgrades. It can be viewed as either an inherent property of the product that facilitates maintenance activities or as the quality-in-use experienced by maintainers when performing maintenance tasks. 
    
\subsubsection{Adaptability} as defined by IEC 25010~\cite{ISO-IEC-25010-2023}, is the capability of a product to be efficiently modified or transferred to different hardware, software, or operational environments. Adaptations may be performed by specialized support staff, operational staff, business users, or end users. When end users are responsible for making adaptations, adaptability aligns with the concept of suitability for individualization as defined in ISO 9241-110. Challenges arising from differences or changes in the actual context of use, beyond those initially specified, can be addressed by iterating the quality improvement cycle to re-evaluate requirements and resolve issues. Enhancing adaptability and flexibility often involves leveraging related sub-characteristics, such as inclusivity and user assistance. Examples of such contexts include using the system in unintended environments (e.g., underwater or in outer space), operating in constrained conditions (e.g., with limited energy or no network connectivity), or involving users or stakeholders who were not initially considered. 
    
\subsubsection{Reusability} is the capability of a product to be used as assets in more than one system or to build other assets~\cite{ISO-IEC-25010-2023}.

As stated earlier in the case of iCAS, operational concerns are more important compared to evolutionary concerns. This is due to the nature of the iCAS business domain. To elucidate this concept, let us consider the example of an oil refinery or a petrochemical production facility. Over the years, the operational procedures within oil and/or petrochemical production lines exhibit minimal change requirements. This reasoning is also applicable to control systems that oversee and regulate aircraft landing and take-off operations or baggage handling systems within airports. These systems are subject to a minimal change requirement and often operate for many years without the need for significant change\footnote{Note that we do differentiate between changing and reconfiguring. In the context of this thesis, changeability does not excel in reconfigurability.}. 

In the context of manufacturing systems, the predominant factor is the flexibility that arises from the system's ability to be reconfigured (i.e., behavioral changeability), an aspect that has been extensively addressed in the literature, particularly in the work of~\cite{ElMaraghy-2005} not the changeability which we discuss above. Certainly, this does not mean that we should exclude considerations of evolutionary concerns entirely from our design equation. Instead, it implies that we must seek an appropriate equilibrium among these concerns.

\subsection{Concerns Tension Diagram}
While operational concerns are complementary, evolutionary concerns often conflict, making decoupled design essential and complex to effectively address iCAS design concerns. For example, reusability and changeability are conflicting design concerns. Reusability accelerates design and development by leveraging reusable elements (e.g., function blocks), but it also introduces coupling, reducing changeability due to cascading changes.

\begin{figure}[h]
    \centering
    \includegraphics[width=0.7\linewidth]{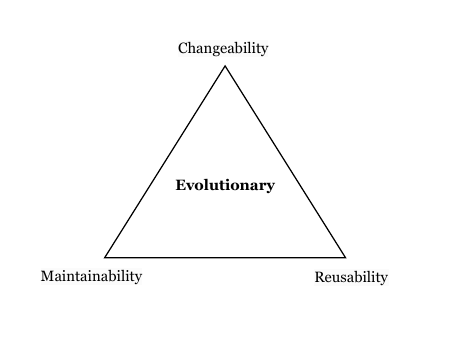}
    \caption{Evolutionary Concerns Tension Diagram}
    \label{fig:concerns_tension_diagram}
\end{figure}

\section{Design Language}
\label{design_language}
Another frequently occurring problem that is not specific to the field of iCAS is the interchangeable use of \textit{design} and \textit{design language}, despite being distinct concepts, yet interrelated. Not distinguishing between them has led to various works~\cite{Sonnleithner-2021,Zhabelova-Vyatkin-2015,Thramboulidis-2012,Werner-2009,Witsch-2009} that attempt to solve design challenges by refining the \textit{design language}. 

Similarly to other design languages that support design frameworks, such as Object-Oriented Languages (e.g., C++, JAVA), which underpins Object-Oriented Design~\cite{Larman-2005,Booch-2007}, the IEC 61131-3 and IEC 61499 should also be recognized as \textit{design languages} rather than \textit{design frameworks}. Each of these languages is addressing specific business needs, and we indeed should not compare them or prefer one to another, but rather we could use each language to address different needs in iCAS. A detailed study by Gsellmann as his thesis~\cite{Gsellmann-2018} and others~\cite{Thramboulidis-2013b} supports this conclusion.

Note that, while a well-structured design language is essential for enabling a robust design framework, improving the language alone cannot resolve fundamental flaws in the framework itself. In the best case, it simply shifts the problem elsewhere, leading to an endless cycle of optimization.

\section*{Acknowledgment}
I would like to express my sincere gratitude to Professor Nam P. Suh for his invaluable support and guidance, which significantly deepened my understanding of design theory.

\section{Conclusion} \label{conclusion}
This study clarifies fundamental concepts of design in the software industry by addressing misconceptions and aligning definitions with advancements in design theory. By examining the nature and beginning of design, distinguishing between design and design language, and contrasting ad-hoc with systematic approaches, we highlight the importance of scientifically grounded design practices. Furthermore, we emphasize the need to balance operational needs with evolutionary challenges to achieve robust and adaptable design outcomes.

Future research should build on this foundation to further refine the application of design theory in industrial software systems and develop tools that actively support the systematic decomposition, validation, and evaluation of functional requirements throughout the design lifecycle. This will allow the iCAS community to move from intuition-driven to axiom-driven practices—achieving more robust, adaptable, and maintainable systems in the decades to come. 

We encourage iCAS scholars and practitioners to utilize the stipulated design definition as a foundational reference prior to engaging in discussions—such as debating the appropriateness of IEC 61131-3 or IEC 61499, both of which cater to distinct requirements. It is imperative to first elucidate the business needs under investigation, and subsequently, apply the axioms introduced in this study to assess which tool offers superior support, rather than resorting to an ad hoc and subjective comparison. 
\bibliography{references}
\bibliographystyle{IEEEtran}

\end{document}